\documentclass[aip,jap,preprint,nofootinbib]{revtex4-1}
\usepackage{graphics}

\usepackage{graphicx}

\usepackage{amssymb}
\usepackage{siunitx}
\usepackage[utf8]{inputenc}

\begin{document}

\title[Increased magnetocrystalline anisotropy in epitaxial Fe-Co-C thin films with spontaneous strain]{Increased magnetocrystalline anisotropy in epitaxial Fe-Co-C thin films with spontaneous strain}

\author{L. Reichel}
\email{l.reichel@ifw-dresden.de}
\affiliation{IFW Dresden, P.O. Box 270116, 01171 Dresden, Germany}
\affiliation{TU Dresden, Faculty of Mechanical Engineering, Institute of Materials Science, 01062 Dresden, Germany}
\author{G. Giannopoulos}
\affiliation{Demokritos National Center of Scientific Research, 15310 Athens, Greece}
\author{S. Kauffmann-Weiss}
\affiliation{IFW Dresden, P.O. Box 270116, 01171 Dresden, Germany}
\affiliation{TU Dresden, Faculty of Mechanical Engineering, Institute of Materials Science, 01062 Dresden, Germany}
\author{M. Hoffmann}
\affiliation{IFW Dresden, P.O. Box 270116, 01171 Dresden, Germany}
\author{D. Pohl}
\affiliation{IFW Dresden, P.O. Box 270116, 01171 Dresden, Germany}
\author{A. Edström}
\affiliation{Uppsala University, Department of Physics and Astronomy, 75120 Uppsala, Sweden}
\author{S. Oswald}
\affiliation{IFW Dresden, P.O. Box 270116, 01171 Dresden, Germany}
\author{D. Niarchos}
\affiliation{Demokritos National Center of Scientific Research, 15310 Athens, Greece}
\author{J. Rusz}
\affiliation{Uppsala University, Department of Physics and Astronomy, 75120 Uppsala, Sweden}
\author{L. Schultz}
\affiliation{IFW Dresden, P.O. Box 270116, 01171 Dresden, Germany}
\affiliation{TU Dresden, Faculty of Mechanical Engineering, Institute of Materials Science, 01062 Dresden, Germany}
\author{S. Fähler}
\affiliation{IFW Dresden, P.O. Box 270116, 01171 Dresden, Germany}

\date{\today}

\begin{abstract}

Rare earth free alloys are in focus of permanent magnet research since the accessibility of the elements needed for nowadays conventional magnets is limited. Tetragonally strained iron-cobalt (Fe-Co) has attracted large interest as promising candidate due to theoretical calculations. In experiments, however, the applied strain quickly relaxes with increasing film thickness and hampers stabilization of a strong magnetocrystalline anisotropy. In our study we show that already 2\,at\% of carbon substantially reduce the lattice relaxation leading to the formation of a spontaneously strained phase with 3\,\% tetragonal distortion. In these strained (Fe$_{0.4}$Co$_{0.6}$)$_{0.98}$C$_{0.02}$ films, a magnetocrystalline anisotropy above 0.4\,MJ/m$^3$ is observed while the large polarization of \SI{2.1}{T} is maintained. Compared to binary Fe-Co this is a remarkable improvement of the intrinsic magnetic properties. In this paper, we relate our experimental work to theoretical studies of strained Fe-Co-C and find a very good agreement.

\end{abstract}


\maketitle

\section{Introduction}

Permanent magnets are everywhere in our daily lives and will become more important in near future. There is a need for new materials as the abundance of the common rare earth based alloys has been questioned within the last years\,\cite{Coey}. However, competitive alternatives are still missing. The alloy Fe-Co has often been considered a promising candidate for rare earth free permanent magnets\,\cite{Burkert,Neise,Turek}. Its very high intrinsic magnetic moment\,\cite{Victora,James} is one prerequisite. The second requirement for a powerful permanent magnet is a strong magnetic anisotropy. However, this was not reported yet in Fe-Co samples of considerable size. The reason is the cubic symmetry of the alloy’s unit cell in thermodynamic equilibrium. This causes an only weak cubic magnetic anisotropy\,\cite{Shih}. Typically, the magnetocrystalline anisotropy (MCA) is weak in cubic materials since spin-orbit coupling is lower than in uniaxial crystal structures which tend to exhibit strong magnetic anisotropies. Already a decade ago, it was proposed to overcome the challenge of magnetic anisotropy by straining the unit cell\,\cite{Burkert}. The $c/a$ ratio of the unit cell was used to describe the tetragonal strain where $c$ is the length of the strained axis and $a$ the length of the axes perpendicular to the applied strain. Calculations\,\cite{Burkert,Neise} based on Density Functional Theory (DFT) predicted remarkably high MCAs if the unit cell would be tetragonally distorted with $c/a$ between 1.2 and 1.25. The most common approach to strain the unit cell experimentally\,\cite{Andersson,Winkelmann,Luo,Yildiz} is coherent epitaxial growth of thin films on suitable substrates that provide appropriate in-plane lattice parameters~$a$. Assuming constant volume of the Fe-Co unit cell, a strain parallel to the film normal is expected when $a$ is smaller than the equilibrium lattice parameter of Fe-Co. Up to now, tetragonally distorted Fe-Co films with perpendicular easy axis of magnetization could not be produced with film thicknesses exceeding 15 monolayers (ML)\,\cite{Winkelmann,Luo,Yildiz}. Thicker films do not exhibit a perpendicular easy axis since lattice relaxation takes place. After strain relaxation, shape anisotropy exceeds MCA and dominates the magnetic performance causing in-plane easy axes. This thickness dependent spin reorientation transition has been described recently by Kim and Hong\,\cite{Kim}. 

\begin{figure}
\includegraphics[width=0.5\columnwidth]{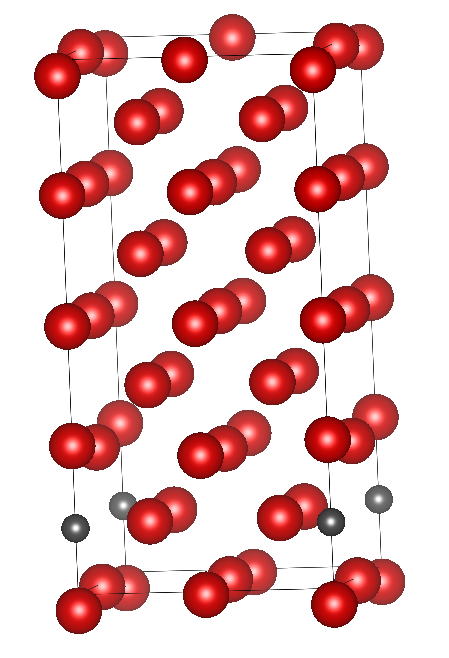}
\caption{\label{DFT}Schematic (Fe$_{0.4}$Co$_{0.6}$)$_{32}$C supercell as applied in the DFT calculations. Fe and Co atoms are marked in red, the smaller C atoms (black) are aligned along the $c$ axis. 
}
\end{figure}

In order to maintain the strain in films thicker than \SI{15}{ML}, new concepts have to be applied. In a recent study, Delczeg-Czirjak et al.\,\cite{Delczeg} calculated that small atoms like carbon may stabilize the strain in Fe-Co. Their DFT calculations identify the preferential positions of the C atoms being interstitials along the $c$ axis of the Fe-Co lattice (Fig.\,\ref{DFT}) preferring lattice positions between two Co atoms. This may be the main reason for the resulting strain since an arbitrary contribution of C atoms would result in an isotropically strained lattice with no tetragonal distortion. Structures of Fe-Co-C phases with up to 11\,at\% carbon have been studied theoretically. The highest magnetic anisotropy energies $MAE$ are reported\,\cite{Delczeg} for (Fe$_x$Co$_{1-x}$)$_{16}$C i.\,e.\,in Fe-Co including 5.9\,at\% C. For such a structure with $c/a = 1.12$, the $MAE$ was calculated as total-energy difference for the two magnetization directions to \SI{0.75}{MJ/m^3}.

Our study presents the first results on Fe-Co thin films alloyed with small amounts of carbon. By measuring the structural film properties during growth, we can compare these Fe-Co-C films to binary Fe-Co films. We show that already 2\,at\%~C stabilizes a tetragonal distortion of the former cubic unit cell. Films with tetragonal distortion of the lattice also show the presence of a MCA pointing in direction of the strained $c$ axis. Their strength is in the range of the predictions from DFT calculations. Since the structural and magnetic properties do not diminish at high film thickness, Fe-Co with 2\,at\%~C is considered to be a spontaneously strained phase.

\section{Experimental and theoretical methods}

Film preparation was performed by Pulsed Laser Deposition (PLD) in ultra high vacuum ($5\times 10^{-9}$\,mbar) at room temperature. A KrF excimer laser (Lambda Physik LPX 305i) with \SI{248}{nm} wavelength and pulse length of \SI{25}{ns} was used. The film substrates were MgO(100) single crystals. Prior the Fe-Co-C layer, a \SI{50}{nm} thick Au-Cu buffer layer was deposited on a \SI{3}{nm} thick Cr seed layer. The composition of the Au$_x$Cu$_{1-x}$ layer was adjusted by co-deposition from Au and Cu element targets, i.\,e.\,the laser pulses hit the targets in an alternating manner. For the Fe-Co-C layers, co-deposition from three targets was applied which were element targets of Fe and Co and a Fe$_{80}$C$_{20}$ composite target. During film growth, in situ reflection high energy electron diffraction (RHEED) was performed to study the lattice properties. The electron energy was \SI{30}{keV}. The diffraction pattern was recorded by a CCD camera which was triggered by the PLD software. Each RHEED record could be linked to a certain film thickness as the deposition rates were prior measured with a rate monitor. With a user developed analysis software, the lattice parameters $a$ and $c$ of the grown film were obtained by fitting the RHEED reflections.

Film compositions were confirmed by energy dispersive x-ray (EDX) measurements on a Bruker EDX at a JEOL JSM6510-NX electron microscope. The carbon content was measured by Auger Electron Spectroscopy (AES) on a JEOL JAMP-9500F Field Emission Auger Microprobe device after sputter cleaning of the surface with Ar$^+$ ions. As reference material for the C quantification the Fe$_{80}$C$_{20}$ target was used. The surface topography was investigated by atomic force microscopy on an Asylum Research Cypher AFM. X-ray diffraction (XRD) was performed on a Bruker D8 Advance diffractometer in Bragg-Brentano geometry (Co-K$\alpha$ radiation). Pole figure measurements were carried out on an X’pert four circle goniometer (Cu-K$\alpha$ radiation). Transmission electron microscopy (TEM) investigations were conducted to confirm the obtained distortion on a Titan$^3$ 80-300 microscope, equipped with an imaging C$_S$ corrector and a Schottky field emission electron source. The film lamellae were cut with focused ion beam on a FEI microscope (Helios NanoLab 600i). Magnetic measurements were performed in a Quantum Design Physical Property Measurement System using a vibrational sample magnetometer (VSM) which operated at \SI{40}{Hz} and \SI{300}{K}. 

Two different density functional theory (DFT) methods were utilized in the computational part of this work. First, full potential code WIEN2k\,\cite{Blaha} with linearized augmented plane waves as basis functions was used. This method treats core electrons fully relativistically, while valence electrons are treated in the scalar relativistic approximation with spin-orbit coupling, which is essential for the calculation of $MAE$, added in a second variational approach\,\cite{Koelling}. The $MAE$ was evaluated using the force theorem\,\cite{Jansen} and disorder was treated by the virtual crystal approximation (VCA) which is, however, known to overestimate $MAE$ in Fe-Co based systems, although it reproduces the correct qualitative behavior\,\cite{Turek,Neise,Delczeg}. Hence, a second DFT approach, namely the spin polarized relativistic Korringa-Kohn-Rostoker (SPR-KKR) code\,\cite{Ebert1,Ebert2} was used. In SPR-KKR all electrons are treated fully relativistically, but a spherical shape approximation is applied for the potential. Disorder is treated via the more realistic coherent potential approximation (CPA) and $MAE$ is evaluated by total energy differences for two different magnetization directions. In both DFT methods the generalized gradient approximation\,\cite{Perdew} was used for the exchange-correlation potential.

\section{In situ observations during Fe-Co(-C) film growth}

\begin{figure}
\includegraphics[width=0.6\columnwidth]{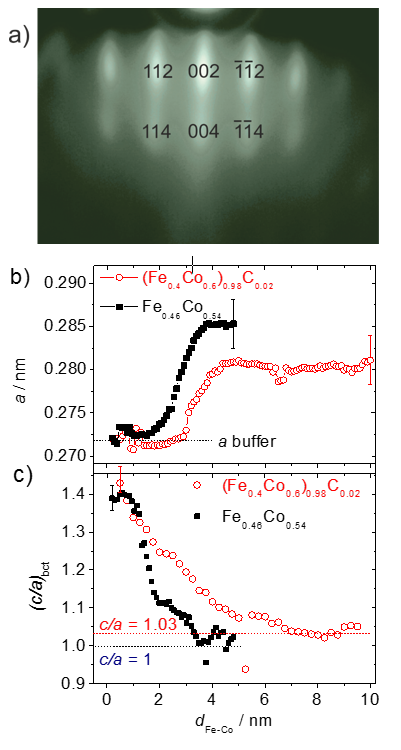}
\caption{\label{RHEED}a) Exemplary RHEED pattern of a (Fe$_{0.4}$Co$_{0.6}$)$_{0.98}$C$_{0.02}$ film of \SI{2.4}{nm} thickness recorded during film deposition along the FeCo[110] azimuth. The horizontal distance between the reflections was used to obtain the reciprocal in-plane lattice parameter\,$a$, the vertical distance gives the out-of-plane lattice parameter\,$c$. b) In-plane lattice parameters\,$a$ of a (Fe$_{0.4}$Co$_{0.6}$)$_{0.98}$C$_{0.02}$ film (open circles) compared to a binary Fe-Co film (full squares). c) $c/a$ ratios for these films.}

\end{figure}

According to theory\,\cite{Burkert,Neise}, the magnetic properties of coherently grown Fe-Co films should depend on the applied in-plane lattice parameter of the buffer layer. As a versatile buffer layer, we selected Au$_x$Cu$_{1-x}$. In this binary system, composition can be used to tune the buffer lattice parameter in a wide range\,\cite{Kauffmann}. For our study, Au contents $x$ between 0.45 and 0.71 were applied which led to $a_{buffer}$ between \SI{0.272}{nm} and \SI{0.281}{nm}\footnote{In difference to all other presented films, the binary film was deposited on an Ir buffer layer. The lattice parameter of Ir ($a_{bcc} = a_{Ir}/\sqrt{2}=0.272$\,nm) is identical when compared to the Au-Cu buffer of the Fe-Co-C film in Fig.\,\ref{RHEED}b and c.}. This is in agreement with studies on sputtered Au$_x$Cu$_{1-x}$ films by Kauffmann-Weiss et al.\,\cite{Kauffmann}, where a small positive deviation from Vegard’s Law for a mixed Au-Cu crystals was reported.

In order to monitor the film growth in situ, RHEED patterns for the [110] azimuthal direction of Fe-Co were recorded during film deposition. An exemplary RHEED pattern for the Fe-Co-C film is shown in Fig.\,\ref{RHEED}a. As measured with atomic force microscopy, the grown films have a roughness RMS of approx. \SI{0.3}{nm}. The film surface thus does not grow in layer-by-layer-mode. Accordingly, a dot-like diffraction pattern is detected which allows for the determination of both lattice parameters $c$ and $a$. An automated analysis, fitting the detected intensity profile with peak functions was used. In Fig. \,\ref{RHEED}b, in-plane lattice parameters $a$ derived from RHEED measurements are plotted versus the thickness of grown film $d_{Fe-Co}$. Fig.\,\ref{RHEED}c depicts the $c/a$ ratio in dependence on $d_{Fe-Co}$. In the graphs, a binary Fe$_{0.46}$Co$_{0.54}$ film (full squares) is compared to a ternary Fe$_{0.4}$Co$_{0.6}$ film containing 2\,at\% C (open circles). 

At first, we will discuss the results for the binary Fe$_{0.46}$Co$_{0.54}$ film. The progression of the in-plane lattice parameter (full squares, Fig.\,\ref{RHEED}b) shows that $a$ first keeps a nearly constant value which is identical to the lattice parameter of the used buffer a$_{buffer}$ in this crystallographic direction. Hence, the first \SI{2}{nm} of the film grow coherently. This is equivalent to \SI{14}{ML} and confirms observations of other groups which studied ultrathin epitaxial Fe-Co films\,\cite{Andersson,Winkelmann,Luo,Yildiz}. They reported that only Fe-Co films with thicknesses up to \SI{15}{ML} exhibit a strong magnetocrystalline anisotropy due to tetragonal distortion. In this paper, however, we describe film thickness in nm units as we intend to increase the thickness where strain is still present. 

With further increasing the film thickness, the described lattice coherence is lost in the Fe$_{0.46}$Co$_{0.54}$ film. This limit is called critical thickness $d_C$. In the following, $a$ increases continuously and, within less than \SI{2}{nm} additionally grown film, reaches a value at which it finally remains. The increase of $a$ reveals the lattice relaxation which commonly proceeds by introduction of misfit dislocations. These dislocations do not form before a critical thickness $d_C$ is reached because of the required dislocation energy. The described relaxation ends when a final value of $a$ is reached, where no more changes in the lattice occur. For the binary film, this final value is equal to the equilibrium lattice parameter\,\cite{Predel}. The described mechanisms are also observed on the $c/a$ curve in Fig.\,\ref{RHEED}c. For the first monolayers of grown film, $(c/a)_{bct}$ is close to $\sqrt{2}$ which is the expected value for an fcc structure described by the Bain transformation using a body centered tetragonal (bct) unit cell\,\cite{Bain}. Having reached $d_C$, the decrease of $(c/a)_{bct}$ becomes stronger which can be attributed to a large number of occurring misfit dislocations. After \SI{2}{nm} additional deposited film i.\,e.\,at $d_{Fe-Co} = 4$\,nm, the relaxation is completed. For the Fe$_{0.46}$Co$_{0.54}$ film (full squares), the final value for $(c/a)_{bct}$ is 1. This is consistent to its bcc equilibrium structure. In additional RHEED studies we observed that an Fe content variation in binary Fe$_y$Co$_{1-y}$ between $y=0.3$ and $y=0.6$ has no influence on its lattice relaxation. This allows a comparison between this film and the ternary (Fe$_{0.4}$Co$_{0.6}$)$_{0.98}$C$_{0.02}$ film which we report in the following. 

From the RHEED derived in-plane lattice parameters of the Fe-Co-C film (open circles, Fig.\,\ref{RHEED}b), we already find important differences when compared to the binary film (full squares). Although lattice relaxation is also observed, the critical thickness where the strain starts to relax is increased to approx. \SI{3}{nm} and also the thickness at which $a$ reaches its final value is slightly increased. Further, the final in-plane lattice parameter after relaxation gives a striking difference. The C containing film finishes its lattice relaxation at a reduced value of $a = 0.281$\,nm. Regarding $c/a$ ratios of the films (Fig.\,\ref{RHEED}c), a peculiar difference is the reduced slope of the curve for the C containing film (open circles) when compared to the binary Fe$_{0.46}$Co$_{0.54}$ film (full squares). The relaxation thus is considerably slower, reaching the final $(c/a)_{bct}$ value after \SI{7}{nm} of grown film. After relaxation $(c/a)_{bct}$ adapts a higher final value for the ternary Fe-Co-C film, in contrast to the binary film where $c/a$ is 1. For the (Fe$_{0.4}$Co$_{0.6}$)$_{0.98}$C$_{0.02}$ film it remains at approx. 1.03. This observation of a residual tetragonallity together with the indication, that the formation of misfit dislocations has a lower driving force, already implies that Fe-Co-C exhibits a spontaneous tetragonal distortion compared to the induced distortion in binary Fe-Co.
The described mechanisms of film growth were already reported in-depth on thin films of pure Fe e.\,g.\,by Roldan Cuenya et al.\,\cite{Roldan}. However, our results aim at the differences between the two regarded films which have very different structural properties after lattice relaxation.

\section{Ex-situ observed structural properties of the films}

\begin{figure}
\includegraphics[width=\columnwidth]{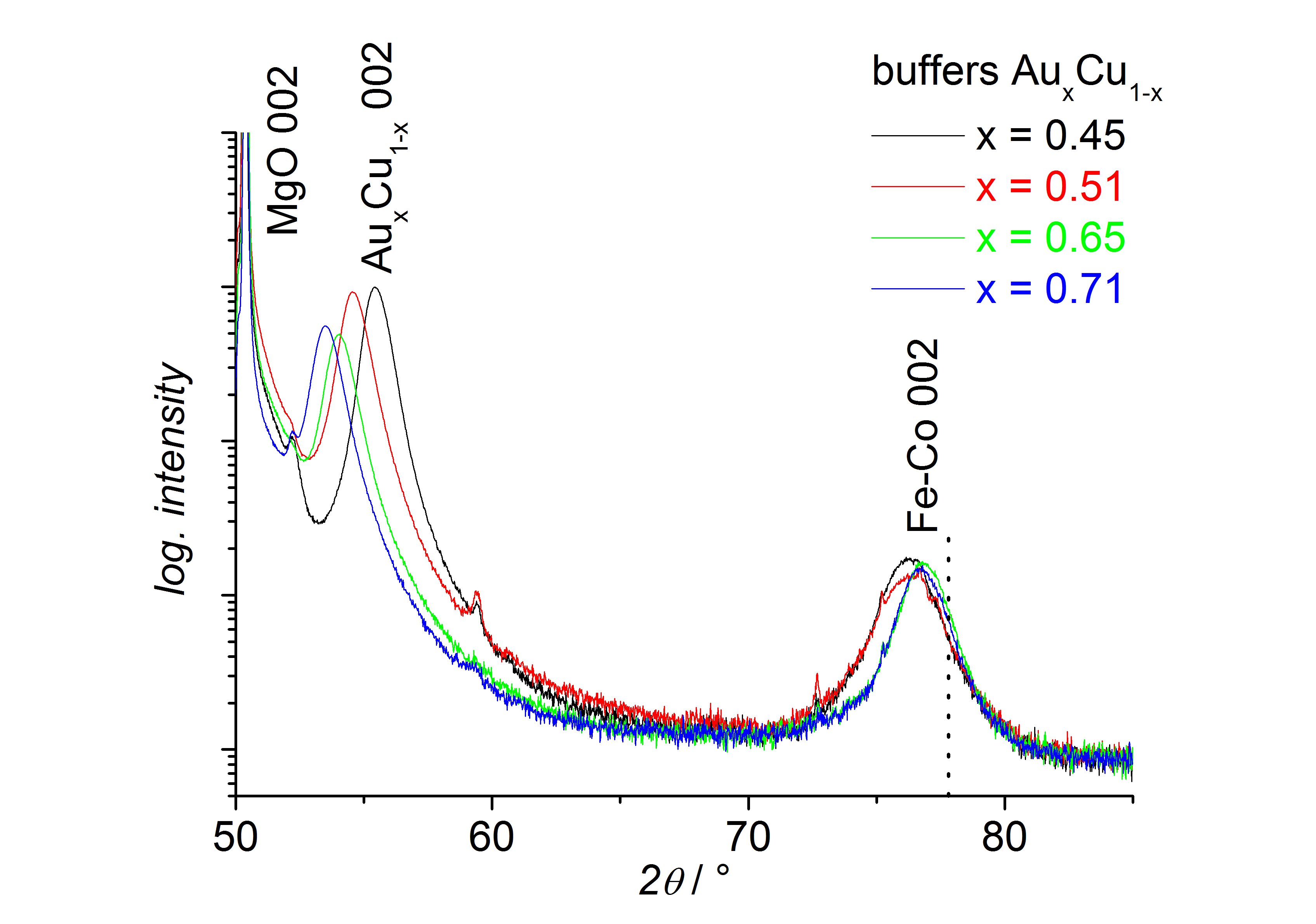}
\caption{\label{XRD}XRD results for (Fe$_{0.4}$Co$_{0.6}$)$_{0.98}$C$_{0.02}$ films on different Au$_x$Cu$_{1-x}$ buffer layers. The equilibrium 002 reflection of binary bcc Fe-Co is marked with a dotted line\,\cite{Predel}. The shift of the 002 reflection to a lower angle indicates a strained $c$ axis in the deposited films. The 002 reflection of the Au-Cu buffer varies between the samples since a different composition implies a changing lattice parameter. The position of the Fe-Co-C 002 reflection is independent on the chosen buffer lattice parameter which varies with the Au-Cu composition. 
}
\end{figure}

For a more detailed examination of the influence of the buffer lattice parameter on the (Fe$_{0.4}$Co$_{0.6}$)$_{0.98}$C$_{0.02}$ films, we increased film thickness to \SI{20}{nm} and applied XRD in Bragg-Brentano geometry. The detected patterns (Fig.\,\ref{XRD}) confirm the change of the $c$ axis length since the 002 reflections of Fe-Co are shifted to lower 2$\Theta$ angles. The expected value for binary bcc Fe$_{0.4}$Co$_{0.6}$ is marked as a dotted line. When comparing films with the same composition deposited on different Au$_x$Cu$_{1-x}$ buffers no variation of the $c$ axis length is observed. Its length is calculated as equal to \SI{2.91}{\AA} which is 2.1\,\% higher than the equilibrium value of cubic Fe$_{0.4}$Co$_{0.6}$\,\cite{Predel}. 

\begin{figure}
\includegraphics[width=\columnwidth]{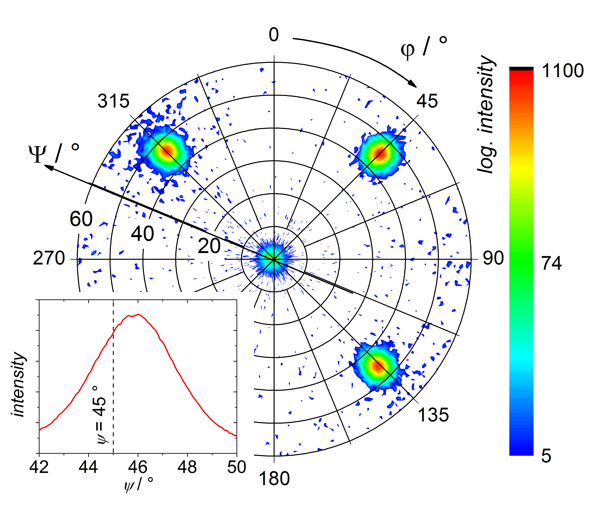}
\caption{\label{texture}XRD 011 pole figure of a \SI{100}{nm} thick (Fe$_{0.4}$Co$_{0.6}$)$_{0.98}$C$_{0.02}$ film on Au$_{0.53}$Cu$_{0.47}$. Its fourfold symmetry proves a square base of the lattice. The positive deviation of $\psi$ from \SI{45}{^{\circ}} (inset) indicates the tetragonal strain which is 1.03 for this sample. The MgO(100) edges [100] and [010] are parallel to the figure edges. 
}
\end{figure}

As independent probe of the tetragonal distortion, we use pole figure measurements. Fig.\,\ref{texture} depicts the 011 pole figure of a \SI{100}{nm} thick (Fe$_{0.4}$Co$_{0.6}$)$_{0.98}$C$_{0.02}$ film. Epitaxial film growth is confirmed by the four intensity maxima. In this pole figure, the tilt angle $\psi$ of the sample, where the 011 pole gets its highest intensity, is a measure for the tetragonallity of the measured lattice. If it is equal to \SI{45}{^{\circ}}, the unit cells are cubic with $c/a$ of 1. This is the case for binary Fe-Co films (not shown). In the present pole figure, $\psi$ has its maximum at \SI{45.9}{^{\circ}} as seen in the inset. With $(c/a)_{bct} = \mathrm{tan} \psi$ the tetragonal distortion of the lattice is determined to 1.03. From this result, we conclude that even thicker films exhibit the same lattice distortion like \SI{10}{nm} thin films. This residual strain of the lattice is a strong indication that Fe-Co-C exhibits a spontaneous strain.  

\begin{figure}
\includegraphics[width=0.7\columnwidth]{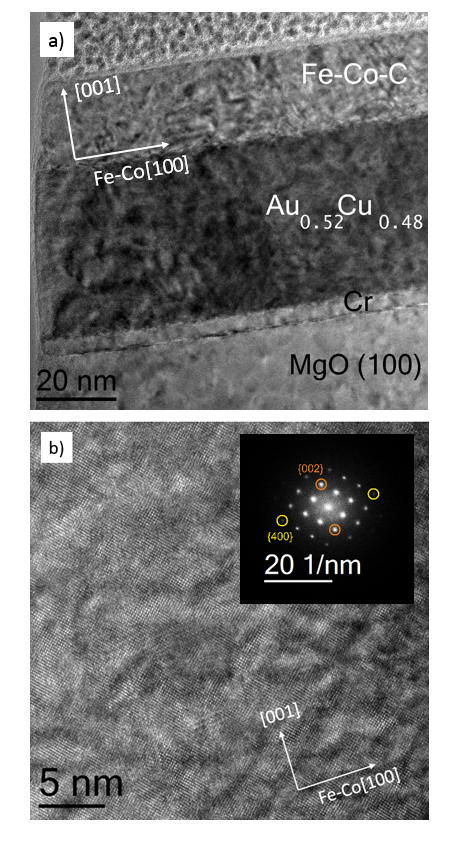}
\caption{\label{TEM}a) TEM image of the complete layers architecture of a 20 nm (Fe$_{0.4}$Co$_{0.6}$)$_{0.95}$C$_{0.05}$ film. b) Magnification of a \SI{100}{nm} (Fe$_{0.4}$Co$_{0.6}$)$_{0.98}$C$_{0.02}$ film oriented in [010] zone axis. The inset depicts the FFT image of this section with the reflections marked which were taken for determination of $a$ (yellow) and $c$ (orange). 
}
\end{figure}

Atomic resolution TEM investigations confirm a tetragonal strain in the Fe-Co-C films. Fast Fourier transforms (FFT) of different film sections allow for the determination of the lattice parameters. Fig.\,\ref{TEM} shows exemplary high resolution TEM images of Fe-Co-C films. In Fig.\,\ref{TEM}a, the cross section of the complete layer structure is depicted confirming the continuity of all layers. Fig.\,\ref{TEM}b shows a section of the \SI{100}{nm} thick (Fe$_{0.4}$Co$_{0.6}$)$_{0.98}$C$_{0.02}$ film. Its inset gives the FFT pattern of this image. By measuring the distances between the intensity maxima in [001] and [100] direction, $(c/a)_{bct}$ was determined to 1.02$\pm$0.02. This value is slightly reduced when compared to the XRD measurements. We attribute this reduction to an additional lattice relaxation which occurs when the film is cut to a lamella. This assumption is based on the observation that thinner sections of the lamellae exhibited lower $c/a$ ratios than thicker sections. FFT patterns of different film sections with varying distance to the buffer interface do hardly reveal any difference in tetragonal distortion. This confirms that the lattice relaxation does not proceed with increasing film thickness, which is in agreement with the XRD measurements. However, the $(c/a)_{bct}$ ratio of the first \SI{4}{nm} of the Fe-Co-C film i.\,e.\,close to the interface to the Au-Cu buffer is 1.04$\pm$0.03 and thus slightly higher than the $(c/a)_{bct}$ ratio measured within the film. At the first glance, this appears to contradict the RHEED measurements, where a substantially higher strain was observed in this thickness range. We attribute this to the fact that RHEED is an in situ method while TEM studies the films ex-situ. We argue that most of the formed misfit dislocations propagate through the already grown film until they reach the interface. This explains, why we do not confirm the much higher values of $(c/a)_{bct}$ as observed with RHEED during film growth. However, the slightly higher values of $(c/a)_{bct}$ at the interface indicate an influence of the buffer in this film section. XRD pole figure measurements of \SI{5}{nm} thick films also reveal $(c/a)_{bct}$ of 1.05$\pm$0.01 (not shown) and confirm this tendency. 

As described in the experimental section, the Fe-Co-C films were prepared by co-deposition from different targets. Since one of the used targets contained 20\,at\%~C, we could study the influence of the C content on the structural film properties. AES measurements of additionally prepared films reveal that the highest achieved C content in continuous Fe-Co-C films was 5\,at\%. A further increase of C content leads to a partial delamination of the films. However, the comparison of (Fe$_{0.4}$Co$_{0.6}$)$_{0.98}$C$_{0.02}$ films and films with up to 5\,at\%~C, does not indicate structural differences -- neither in XRD (not shown) nor in TEM. The determined lattice parameters remain unaltered as already described and $(c/a)_{bct}$ remains 1.03. The only difference is a reduced X-ray coherence length in the films with higher C content which indicates a reduction of crystal size. Since the lattice does not change with increasing C content, we propose that the maximum solubility of C in Fe-Co films is 2\,at\%~C. We argue that additional C dissolves and accumulates at grain boundaries which finally, at a C content above 5\,at\%, destabilizes the films. Compared to its maximum solubility in bcc equilibrium Fe-Co of only 0.3\,at\%\,\cite{MSIT}, we can reach significantly higher C contents in our films. We attribute this to the high energy of the deposited ions in PLD\,\cite{Faehler} which allows for a preparation of supersolutions with substantially higher solubility limits than in equilibrium\,\cite{Krebs}. 

\section{Magnetic properties of the strained films}

\begin{figure}
\includegraphics[width=0.8\columnwidth]{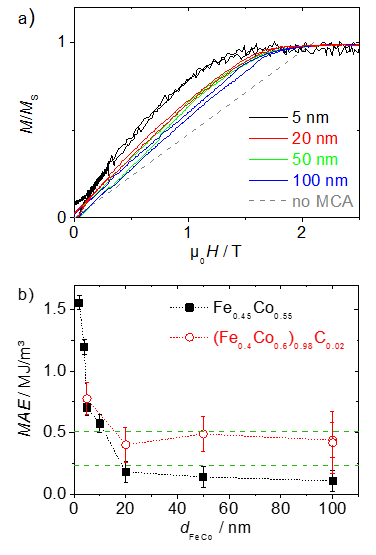}
\caption{\label{magnetic}a) First quadrant of hysteresis curves of (Fe$_{0.4}$Co$_{0.6}$)$_{0.98}$C$_{0.02}$ films with different thicknesses $d_{Fe-Co}$. For comparison, the expected out-of-plane curve for a film with shape anisotropy only (grey broken line) has been added. b) Magnetic anisotropy energies of these films (open circles) compared to binary Fe-Co films (full squares). The broken horizontal lines mark the range of the DFT results of (Fe$_{0.4}$Co$_{0.6}$)$_{32}$C with $c/a = 1.03$ which were obtained by applying two different methods as described in the text.
}
\end{figure}

In order to investigate the influence of the tetragonal lattice distortion on the magnetic properties, VSM measurements were performed at room temperature. Fig.\,\ref{magnetic}a depicts the first quadrants of the out-of-plane (oop) hysteresis curves of (Fe$_{0.4}$Co$_{0.6}$)$_{0.98}$C$_{0.02}$ films with different thicknesses on Au$_{0.5}$Cu$_{0.5}$ buffers. The magnetic saturation $\mu_0M_S$ of all of these films was determined to be $2.1\pm 0.1$\,T which agrees with DFT as discussed later. To improve the comparability of the hystereses, the maximum magnetizations were normalized. All films saturate at a magnetic field $\mu_0H_S$ which is lower than \SI{2.1}{T}. This implies the presence of an oop component of magnetic anisotropy competing with shape anisotropy for all film thicknesses. For comparison, a hypothetical curve of a film with only shape anisotropy and no oop component of magnetic anisotropy (grey broken line) has been added. As the epitaxial growth results in a unique alignment of all tetragonal unit cells with their $c$ axis perpendicular to the substrate, we can attribute this additional anisotropy in the (Fe$_{0.4}$Co$_{0.6}$)$_{0.98}$C$_{0.02}$ films to the tetragonal distortion of the lattice which makes it a MCA. As seen in Fig.\,\ref{magnetic}a, the film thickness has an influence on the slope of the hystereses and, therefore, also on its MCA. The film with the lowest thickness, \SI{5}{nm}, shows the highest slope of its hysteresis i.\,e.\,the highest MCA. This is consistent to the texture measurements, where for the thinnest film, a slightly higher tetragonal distortion is measured. Also TEM revealed a slightly increased distortion of the first nanometers of grown film. For thicker films, the shape of the hysteresis curves is unaltered (Fig.\,\ref{magnetic}a), i.\,e., there is little change of the magnetic properties. This also matches to the structural observations, where no major difference of lattice distortion was observed for these films. 

From the magnetic hysteresis curves, values for the magnetocrystalline anisotropy energy MAE of the films were estimated. We use the triangle, set in the oop hysteresis curve between (0,0), (0,$\mu_0M_S$) and ($\mu_0H_S$,$\mu_0M_S$) as measure for the effective anisotropy energy $MA_{eff}$. The difference between the shape anisotropy energy $MA_{shape} = 1/2 \cdot \mu_0M_S^2$ and $MA_{eff}$ gives then the $MAE$ of the film. The results of this estimation are plotted in Fig.\,\ref{magnetic}b compared to the $MAE$ of binary Fe-Co films which were determined similarly from their hysteresis curves. For both, the binary and the ternary Fe-Co-C films, the highest $MAE$ is reached for ultrathin films with thicknesses below \SI{10}{nm}. This may be attributed to the larger contribution of the higher strained Fe-Co lattice at the interface. Fe-Co films with oop magnetic easy axis were already reported for thicknesses up to \SI{15}{ML}, i.\,e.\,, \SI{2}{nm}\,\cite{Winkelmann,Luo,Yildiz}. However, the large $MAE$ in the ultrathin films may also originate from surface dipole anisotropies that may benefit magnetic moments to align in perpendicular direction\,\cite{Johnson}. The higher the film thickness, the lower is the contribution of both surface related effects and thus $MAE$ decreases. With increasing film thickness, the $MAE$ of the (Fe$_{0.4}$Co$_{0.6}$)$_{0.98}$C$_{0.02}$ films does not decrease as strongly as it is observed for binary Fe-Co films. While in the latter, only the first few monolayers (see Fig.\,\ref{RHEED}c) are strained tetragonally, the Fe-Co-C lattice exhibits a strain up to film thicknesses of at least \SI{100}{nm}. This is the reason why the $MAE$ of these films is maintained above \SI{0.4}{MJ/m^3}. In the binary films, the fraction of distorted Fe-Co becomes negligible at higher film thicknesses. 

Regarding films with higher C content than 2\,at\%, no further increase of $MAE$ was observed. Instead, a reduction of $\mu_0M_S$ below \SI{2}{T} for 5\,at\%~C was measured, which is a deterioration of magnetic properties compared to the (Fe$_{0.4}$Co$_{0.6}$)$_{0.98}$C$_{0.02}$ films. Theory agrees at this point. More C in the lattice not only reduces the magnetic saturation due to higher fraction of diamagnetic C, but also reduces the magnetic moment of the Fe and Co atoms\,\cite{Delczeg}.  

Our aim is to compare the described experimental observations to theoretical results based on DFT. As already discussed in the introduction, the tetragonal distortion was predicted by Delczeg-Czirjak et al.\,\cite{Delczeg} for Fe-Co alloyed with carbon. The present study proves these findings also for lower C contents of only 2\,at\%. However, higher C contents did not further strain the unit cell as predicted in the literature\,\cite{Delczeg} where 6\,at\% C led to $(c/a)_{bct}$ ratios of 1.12. We suggest a thermodynamic limit of solubility of C as most likely reason. Additional DFT calculations were performed based on the structural observations of our (Fe$_{0.4}$Co$_{0.6}$)$_{0.98}$C$_{0.02}$ films. The input parameters were $c = 0.290$\,nm and $a = 0.281$\,nm yielding a $(c/a)_{bct}$ ratio of 1.03 as experimentally observed with XRD (Fig.\,\ref{XRD} and \ref{texture}). This $(c/a)_{bct}$ ratio is similar to that of $(c/a)_{bct} = 1.033$ theoretically predicted\,\cite{Delczeg} for (Fe$_{0.4}$Co$_{0.6}$)$_{24}$C. The here applied film composition was (Fe$_{0.4}$Co$_{0.6}$)$_{32}$C i.\,e.\,3\,at\% C as depicted in Fig.\,\ref{DFT}. The internal positions of atoms were relaxed in WIEN2k after which the same structure was used as input to SPR-KKR. Evaluation of the MCA with 1200\,\textbf{k} vectors and smallest muffin-tin radius times maximum \textbf{k} vector used in-plane wave basis functions chosen to $RK_{max} = 8.3$ yielded $MAE = 0.51$\,MJ/m$^3$ and $\mu_0M_S = 2.1$\,T. When MCA is determined in SPR-KKR, with CPA and 2600\,\textbf{k} vectors, instead of VCA in Wien2k, it is reduced to $MAE = 0.224$\,MJ/m$^3$, but $\mu_0M_S$ is still \SI{2.1}{T}. Comparing the results from both applied DFT approaches we find a good agreement with our experimental data e.\,g.\,$MAE = 0.44\pm 0.14$\,MJ/m$^3$ for the \SI{100}{nm} thick (Fe$_{0.4}$Co$_{0.6}$)$_{0.98}$C$_{0.02}$ film. As seen in Fig.\,\ref{magnetic}b, the two calculated $MAE$ could be taken as upper and lower limits for the experimentally obtained anisotropy energy for film thicknesses from \SI{20}{nm} upwards. However, the experimental $MAE$ seems to be closer to the VCA based theoretical value than to the CPA based one. Already Delczeg et al.\,\cite{Delczeg} reported, that CPA might underestimate the effect of disorder on MCA when compared to more accurate special quasirandom structures (SQS). 

The observed good agreement between experimental and theoretical results also indicates that a reduction of C content from 3 to 2\,at\% does not lower the MCA when the state of strain is kept. As expected, the MCA is more sensitive to the $c/a$ ratio than to the C content. An additional explanation is that a higher amount of C atoms, which is expected to result in a higher strain\,\cite{Delczeg}, may introduce local disorder. This disorder could cause a reduction of $MAE$ when compared to a highly symmetric distorted Fe-Co crystal with no C atoms\,\cite{Delczeg}. Hence, a not too high C content is expected to be beneficial for structural and magnetic properties.

\section{Conclusion}

In this study, carbon was alloyed to Fe-Co in order to stabilise the tetragonal distortion of the unit cell. It was shown that already a very low fraction of C can benefit the formation of a phase with tetragonal lattice symmetry. For 2 at\%~C in Fe$_{0.4}$Co$_{0.6}$ the former cubic lattice is strained by 3\,\%. This strain is small when compared to theoretical calculations for Fe-Co with higher C content\,\cite{Delczeg}. However, it has a remarkable influence on the magnetic properties as it establishes a uniaxial MCA. Our experimental results confirm the applied theoretical model where the C atoms preferentially occupy positions along the $c$ axis of the Fe-Co lattice. If this did not happen, an isotropically strained cubic lattice would be expected. 
We find strong indications that (Fe$_{0.4}$Co$_{0.6}$)$_{0.98}$C$_{0.02}$ prepared under the named conditions forms a spontaneously strained phase: (1) After an initial lattice relaxation which takes place in the first grown monolayers, the relaxation stops at a state with a lattice distortion of 1.03 (see Fig.\,\ref{RHEED}c). (2) This lattice distortion is independent of film thickness, as it is observed in films with thicknesses between \SI{20}{nm} and \SI{100}{nm}. We find no indications for a driving force that promotes a further relaxation.

The magnetocrystalline anisotropy energy of the described alloy was determined to be approx. \SI{0.44}{MJ/m^3}. This experimental result agrees with bounds predicted by DFT calculations based on two different methods of treating the chemical disorder.

Our study shows that apart from ultrathin films, also thicker films based on Fe-Co can exhibit magnetocrystalline anisotropy. Although their anisotropy is still too low to be competitive to the common hard magnetic alloys, we consider our results as one step towards possible permanent magnet alternatives based on 3d elements. The capability of small atoms like C to strain the former cubic lattice of Fe-Co is a key message of this work. Comparable elements~(B,N) are expected to act similarly. One important advantage of additions of these elements in small amount is that a high magnetic moment is maintained. As a spontaneously strained phase in Fe-Co-C is formed, this approach opens many possibilities that may lead to novel hard magnetic materials. These potential new permanent magnets would not be restricted to thin films.

\acknowledgements
We acknowledge funding of the EU through FP7-REFREEPERMAG. For experimental support, we gratefully thank Ruben Hühne, Steffi Kaschube, Christine Damm and Juliane Scheiter.

\end{document}